\documentclass[iop]{emulateapj}

\newcommand{\degrees}{\mbox{$^{\circ}$}} 
\def\sqig{$\sim$}
\def\lsi{LS\,I\,+61\degrees 303}
\def\ls5039{LS\,5039}

\def\cxou{CXOGSG J140514.4-611827}
\def\J1018{1FGL\,J1018.6-5856}
\def\j1018{1FGL\,J1018.6-5856}
\def\hessj{HESS\,J0632+057}

\def\ergs{erg~s$^{-1}$}
\def\ergcm2s{erg\,cm$^{-2}$\,s$^{-1}$}

\def\Fermi{\textit {Fermi}}
\def\Swift{{\it Swift}}

\def\INTEGRAL{{\it INTEGRAL}}
\def\src{4FGL\,J1405.1-6119}
\def\three_src{3FGL\,J1405.4-6119}
\def\flsrc{FL8Y\,J1405.3-6119}

\def\bfa{}
\def\bfb{}
\def\bfc{}
\def\bfd{}

\def\mybf{}

\usepackage{draftcopy}

\draftcopyName{}{24}

\begin{document}
\def\subtitle{}
\submitted{}
\accepted{August 22, 2019}
\journalinfo{}

\title{Discovery of the Galactic High-Mass Gamma-ray Binary 4FGL J1405.1-6119}

\author{R.~H.~D. Corbet\altaffilmark{1}, L. Chomiuk\altaffilmark{2}, M.~J. Coe\altaffilmark{3}, 
J.~B. Coley\altaffilmark{4}, G. Dubus\altaffilmark{5}, \\
P.~G. Edwards\altaffilmark{6}, P. Martin\altaffilmark{7}, V.~A. McBride\altaffilmark{9}, 
J. Stevens\altaffilmark{6}, J. Strader\altaffilmark{2}, L.~J. Townsend\altaffilmark{8}
}

\altaffiltext{1} {University of Maryland, Baltimore County, and
X-ray Astrophysics Laboratory, Code 662 NASA Goddard Space Flight Center, Greenbelt Rd., MD 20771, USA.
Maryland Institute College of Art, 1300 W Mt Royal Ave, Baltimore, MD 21217, USA.}

\altaffiltext{2} {Department of Physics and Astronomy, Michigan State University, East Lansing, MI 48824, USA.}

\altaffiltext{3} {{\bfa School of Physics and Astronomy}, University of Southampton, Southampton SO17 1BJ, UK.}

\altaffiltext{4} {Department of Physics and Astronomy, Howard University, Washington, DC 20059, USA. 
CRESST/Code 661 Astroparticle Physics Laboratory, NASA Goddard Space Flight Center, Greenbelt Rd., MD 20771, USA.}

\altaffiltext{5} {Institut de Plan\'{e}tologie et d'Astrophysique de Grenoble, Univ. Grenoble Alpes, CNRS, F-38000 Grenoble, France.}

\altaffiltext{6} {Commonwealth Scientific and Industrial Research Organisation Astronomy and Space Science, PO Box 76, Epping, New South Wales 1710, Australia.}

\altaffiltext{7} {Institut de Recherche en Astrophysique et Plan\'etologie, Universit\'e de Toulouse, CNRS, F-31028 Toulouse cedex 4, France.}

\altaffiltext{8} {Department of Astronomy, University of Cape Town, Private Bag X3, Rondebosch, 7701, South Africa.}

\altaffiltext{9} {South African Astronomical Observatory, PO Box 9, Observatory, 7935, South Africa.}

\begin{abstract}
We report the identification from multi-wavelength observations
of the \Fermi\ Large Area Telescope (LAT) source \src\
(= \three_src)
as a high-mass gamma-ray binary.
Observations with the LAT show that
gamma-ray emission from the system is modulated at a
period of 13.7135 $\pm$ 0.0019 days, with the presence of two maxima
per orbit with different spectral properties.
X-ray observations using the {\it Neil Gehrels Swift
Observatory} X-ray Telescope (XRT) show that X-ray emission is also
modulated at this period, but with a single maximum
that is closer to the secondary lower-energy gamma-ray maximum.
A radio source, coincident with the X-ray source
is also found from Australia Telescope Compact Array (ATCA)
observations, and the radio emission is modulated
on the gamma-ray period with similar phasing to the X-ray emission.
A large degree of interstellar obscuration
severely hampers optical observations, but
a near-infrared counterpart is found.
Near-infrared spectroscopy indicates an O6 III spectral
classification.
This is the third gamma-ray binary to be discovered with the Fermi LAT
from periodic modulation of the gamma-ray emission, the other two sources
also {\bfa have} early O star, rather than Be star, counterparts.
We consider at what distances
we can detect such modulated gamma-ray emission with the LAT, and examine
constraints on the gamma-ray binary population of the Milky Way.

\end{abstract}

\section{Introduction}

High-mass gamma-ray binaries (HMGBs) are very rare objects.
They consist of an OB star in orbit with a
compact object, where interactions between
the two objects cause emission whose spectral
energy distribution (SED) peaks above 1 MeV 
\citep[e.g.][and references therein]{Dubus2015a,Dubus2017}.
The principal emission mechanism is thought to be
interactions between the wind of a rapidly rotating
neutron star and the wind from the OB companion.
However, while the presence of a rapidly rotating neutron
star is suspected, it has only been directly detected
in {\bfa the} PSR B1259-63 and PSR J2032+4127 {\bfa binary systems} where the gamma-ray
binary phase is confined to orbital phases near periastron
passage of these highly-eccentric systems with long
orbital periods \citep[e.g.][and references therein]{Li2017,Johnson2018,Abeysekara2018}.
Gamma-ray binaries are expected to
be precursors to High-Mass X-ray Binaries,
which they {\bfa can} evolve into after the neutron star
has spun down sufficiently to allow accretion to occur.
Systems that contain jet-producing accreting black
holes may also be gamma-ray sources. However, in this
case the SED will peak at X-ray energies and these systems can be
classified as ``gamma-ray emitting X-ray binaries''.
Such systems would include Cygnus X-3 \citep[e.g.][]{Corbel2012,Zdziarski2018}
and Cygnus X-1 \citep[e.g.][]{Bodaghee2013,Zanin2016,Zdziarski2017}.

Generally, HMGBs may be divided into those
systems that contain a Be star, 
and those that contain an O star. Be star systems differ
from the O star systems in that Be stars
posses circumstellar disks, that may be transient.
The known O-star HMGBs are LS 5039, 1FGL J1018.6-5856, and LMC P3, while
the known Be-star gamma-ray binaries 
are \lsi, \hessj, {\bfa PSR} B1259-63 and PSR J2032+4127 \citep[e.g.][and references therein]{Dubus2017}.
Note that \hessj, while detectable at \sqig TeV
energies, appears only weakly detected at \sqig GeV energies \citep{Malyshev2016,Li2017},
and for PSR J2032+4127, while the pulsations away from
periastron were discovered at \sqig GeV energies with the LAT \citep{Abdo2009}, 
binary related activity was primarily detected at
TeV rather than GeV energies \citep{Abeysekara2018}.

The number of detectable Galactic gamma-ray binaries
will depend on factors that include the distribution of
initial neutron star spin periods, and the lifetimes
of gamma-ray binaries.
Initial estimates of the possible population of
gamma-ray binaries came from \citet{Meurs1989} who predicted
\sqig30 systems in the appropriate evolutionary stage.
More recently, based on light-curve modeling and the currently known
number of systems, \citet{Dubus2017}
estimated that \sqig100 Galactic systems may be detectable. 
Due to the large point spread function of gamma-ray telescopes such
as the Fermi LAT \citep[e.g.][]{FermiLAT2019}, there are very
many known gamma-ray sources whose nature is not yet known, and
so are potentially binary systems.

Our program to discover gamma-ray binaries initially
searches for signs of periodic modulation in gamma-ray light
curves obtained with the \Fermi-LAT. We then search for
counterparts at other wavelengths, and for confirmation
of modulation on the same period in the counterparts.
From this program we previously identified \J1018\ and LMC P3
as high-mass gamma-ray binaries. Here we present the {\bfa discovery}
of \src\ as an additional gamma-ray binary.

We present previous observations and analysis of \src\ in
Section \ref{the_source}. We
describe LAT observations and our program to search for modulated
gamma-ray sources in Section \ref{sect:gray_obs}. The \Swift\ X-ray observations
of the region are presented in Section \ref{sect:xray_obs}, and the ATCA observations
in Section \ref{sect:radio_obs}. Near-infrared spectroscopy  is described in Section \ref{sect:ir_obs}. 
The discovery of periodic gamma-ray emission from the direction of \src\ is presented in
Section \ref{sect:gray_results}, the identification of the counterpart as \cxou\ from the
detection of modulated X-ray and radio emission is
given in Sections \ref{sect:xray_results} and \ref{sect:radio_results} respectively.
The nature of \src\ and
the implications for the overall population of gamma-ray binaries in the Galaxy
are discussed in Section \ref{sect:discuss} with an overall conclusion in Section \ref{sect:conclude}.
Unless otherwise stated, uncertainties are given at the 1$\sigma$ level.

\section{Previous Observations of \src\label{the_source}}

\src\ is in the fourth LAT catalog \citep{FermiLAT2019} and
counterparts were also present in previous LAT catalogs
as 1FGL J1405.1-6123c \citep{Abdo2010}, 2FGL J1405.5-6121 \citep{Nolan2012}, \three_src \citep{Acero2015} and
was present in the LAT eight year source list
as \flsrc. In the third catalog of Fermi
sources detected in the 10 GeV to 2 TeV energy range,
it is identified as 3FHL J1405.1-6118 \citep{Ajello2017}.

\citet{Lee2012} examined sources in the second
Fermi LAT catalog to search for
pulsars and identified 2FGL J1405.5-6121
as a candidate based on variability and spectral criteria.
\citet{Saz2016} also undertook a classification of sources in
the 3FGL catalog into pulsars and active galactic nuclei - the two main categories
of identified LAT sources. From this analysis they found that
\three_src\ was most likely to be a pulsar. \citet{Saz2016} also noted
the presence of XRT and Chandra sources {\bfa in the Fermi error region}, 
{\bfa
with the Chandra
source located at 
R.A. = 14$^{\rm h}$05$^{\rm m}$14$^{\rm s}_.$47 
decl. = -61$\degr$18$\arcmin$27$\arcsec_.$7
($\pm$ 0$\arcsec_.19$ statistical, $\pm$0$\arcsec_.8$ systematic).
\citet{Saz2016} also noted
a possible coincidence with the
supernova remnant (SNR) G311.5-00.3 (there listed as ``G311.5+0.3'').}
\three_src\ was also listed as a strong pulsar candidate by
\citet{Wu2018}. 
{\bfa \citet{Clark2017} included \three_src\ in a search
for pulsations from LAT pulsar candidates and obtained a 95\% upper
limit on the pulsed fraction of 0.58.}

Using IBIS on board \INTEGRAL\ covering 2 - 200 keV,
\citet{Landi2017} reported activity from IGR {\bfa J14059-6116}
between 2003 December 7 to 2009 December 14 (MJD 52,980.45 -
55,179.04) {\bfa and proposed that this IBIS source was the counterpart of \three_src.
\citet{Landi2017} also reported on XRT observations which showed
the detection of one source within the IBIS error region
{\bfa
at R.A. = 14$^{\rm h}$05$^{\rm m}$13$^{\rm s}_.$93
decl. = -61$\degr$18$\arcmin$29$\arcsec_.$62 with a 5$\arcsec_.$2 (90\% confidence) error radius.
}
The XRT source was also positionally coincident with}
the near-infrared source 2MASS J14051441$-$6118282/allWISE J140514.40$-$611827.7.
Based on the infrared colors {\bfa of this source},
\citet{Landi2017} considered it unlikely that
\three_src\ was a blazar.

\section{Observations and Analysis}

\subsection{Gamma-ray Observations and Analysis\label{sect:gray_obs}}

The \Fermi\ LAT \citep{Atwood2009} is a pair conversion telescope 
sensitive to gamma-ray photons with energies
between \sqig20 MeV to $>$ 300 GeV.
The LAT data used in this paper were obtained between 2008 August 5 and 2019 March 13 (MJD 54,683
to 58,555). 
During this time, Fermi was primarily operated in a sky survey mode
where the LAT pointing position is alternately rocked away from the zenith to the orbit
north for one spacecraft orbit, then towards the orbit south for
one orbit. In this way, the entire sky is observed every two spacecraft orbits, approximately
every three hours.
For LAT analysis we used the \texttt{fermitools} version \texttt{1.0.1}.
We used the updated Pass 8 LAT data files \citep[``P8R3'',][]{Bruel2018} and the weekly photon files 
provided by the Fermi Science Support Center which include precomputed diffuse response columns.

In our continuing search for new gamma-ray binaries, 
we create light curves for all sources in \Fermi\ LAT catalogs and then calculate power spectra of these to investigate the presence of
periodic modulation.
The third LAT catalog contained 3033 sources \citep{Acero2015}.
The most recent LAT catalog is 4FGL \citep{FermiLAT2019} which contains 5099 sources.
In addition, the eight year source list (``FL8Y''\footnote{https://fermi.gsfc.nasa.gov/ssc/data/access/lat/fl8y/}) 
which was a precursor to the fourth LAT catalog 
contains 5523 sources. In our search for gamma-ray binaries, we analyzed all 3FGL
sources and FL8Y sources. At the time of writing a search of all the sources in
the 4FGL catalog was still in progress. However, after identifying \three_src/FL8Y 1405.3-6119 as
a binary \citep{Corbet2019}, we then used the parameters of this source 
from the 4FGL catalog for a refined analysis of \src.

Light curves covering an energy range of 100 MeV to 500 GeV were
created using a variant of aperture photometry
where, instead of simply summing the number of photons within an
aperture, we estimate the probability that each photon comes from
a source of interest and sum these probabilities \citep[e.g.][]{Kerr2011,FermiLAT2012}. 
To estimate the probability of a photon coming from a source,
models were created for each source using the 3FGL catalog and FL8Y
source list
using sources within a 10 degree radius and \texttt{make3FGLxml} and
\texttt{makeFL8Yxml} respectively. Photon
probabilities were calculated using \texttt{gtsrcprob} and
then summed for a 3 degree radius aperture centered on each source.
We note that although in general the use of ``probability photometry''
increases the signal-to-noise of the light curves, it affects the
photometric properties as probabilities are based on a constant
source brightness. Thus, when a source is brighter than the model
predicts, the probability of a photon coming from the source is underestimated,
and, when the source is fainter than the model, the probability is overestimated. This
results in a decrease of the apparent modulated amplitude.
In addition, there is an energy-dependent effect as the smaller
point spread function (PSF) of the LAT at higher energies results
in higher-energy photons having higher weights.
Time bins of 500s were used for all sources.

Power spectra of these probability-weighted aperture photometry LAT light curves were calculated weighting each
data point's contribution by its relative exposure, after first subtracting
the mean count rate. This is beneficial because of the substantial exposure changes
from time bin to time bin \citep{FermiLAT2009}. For each source the calculated power spectrum covered a period range from 0.05 days
(1.2 hrs) to the length of the light curve, i.e. \sqig 3510 days, giving \sqig70,200 independent
frequencies.
The power spectra were oversampled by a factor of 5 compared to the nominal
resolution, which we take to be the inverse of the length of the
light curve \citep[e.g.][and references therein]{VanderPlas2018}, i.e. \sqig1/3510 days$^{-1}$.
For the strongest peak in each power spectrum
the False Alarm Probability \citep[FAP,][]{Scargle1982}, the estimated probability 
of a signal reaching a power level by chance under the assumption of white noise,
was calculated. This FAP takes into account the number of independent frequencies
searched, but does not include the effect of searching for periodicity in multiple
sources. In addition, possible statistical problems
with the FAP have been noted \citep[e.g.][]{Koen1990,Baluev2008,Suveges2014}.
In our photometric analyses the background is not fitted for each time bin, and artifact signals can be
seen at several periods including \Fermi's \sqig 90 minute orbital period, the survey period at twice this,
one day, the Moon's 27.3 day sidereal period,
the 53 day precession period of the \Fermi\ satellite, and one quarter of a year\footnote{http://fermi.gsfc.nasa.gov/ssc/data/analysis/LAT\_caveats\_temporal.html}.
In addition, because of the broad point-spread function of the LAT, particularly
at lower energies, variability in nearby sources can also cause apparent modulation
in a light curve. Because of these potential artifacts, when apparent evidence of periodic modulation
is found in a LAT light curve, it is highly desirable to be able to confirm the modulation
using observations at other wavelengths.

\subsection{X-ray Observations and Analysis\label{sect:xray_obs}}

{\bfb
The \textsl{Swift} XRT \citep{Burrows2005} is a Wolter I X-ray imaging telescope 
sensitive to X-rays ranging from 0.3 to 10 keV.
Although some previous XRT observations {\bfa of the region} had previously been made dating
back to October 2011,
these were infrequent, had short observation durations, and had been made over a long interval.
We therefore requested additional \Swift\ XRT {\bfa Target of Opportunity (TOO)} observations to cover
two orbital cycles more intensively and with longer durations.
The XRT TOO observations of \src\ {\bfa (\cxou)} took place from 2018 May 16 to June {\bfb 17} {\bfa (MJD\,58,254--MJD\,58,286)}
with exposures ranging from $\sim$3.2\,ks to $\sim$4.0\,ks.  
{\bfa
For completeness, we additionally analyzed the 15 short archival XRT observations, which were performed between 2011 October 4 and 2017 December 13
 (MJD\,55,838 and MJD\,58,100). The exposures of these archival XRT observations ranged from $\sim$90\,s to $\sim$4.6\,ks.
}
}

{\bfb
\src\ was observed in Photon Counting \citep[PC;][]{Hill2004} mode with a readout time of 2.5\,s adopting the standard grade filtering (0--12 for PC).
We reduced and analyzed the data using the Swift XRT product generator
\citep{Evans2007} and the HEAsoft v.6.20 package and calibration files dated 2017 May 1, following
the procedures defined in the XRT Data Reduction Guide \citep{XRTGuide}. The data were
reprocessed with the XRTDAS standard data pipeline package \texttt{xrtpipeline} using the standard
filtering procedure to apply the newest calibration and default screening criteria.
}
 

{\bfa
We find the background-subtracted count rates of our TOO observations to be between 
1.0$^{+1.8}_{-0.9}$$\times$10$^{-3}$ and 7\,$\pm$\,2$\times$10$^{-3}${\bfc \,counts s$^{-1}$}.  
Since our observations of \src\ were not affected by pile-up, we extracted the source spectra from count-dependent circular regions generated by the \textsl{Swift} XRT product generator. The ancillary response files, accounting for vignetting, point-spread function correction, and different extraction regions, were generated and corrected for exposure using the FTOOL packages \texttt{xrtmkarf} and \texttt{xrtexpomap}, respectively.  Due to the large neutral hydrogen column density, the source was not detected at energies below 2\,keV.  We, therefore, restrict our spectral analysis to energies above 2\,keV.

Individual spectra were not useful for analysis, as each spectrum was found to have between 0--21\,counts.  A cumulative spectrum was therefore extracted, which has a total of $\sim$165\,counts, and the total exposure is $\sim$34.9\,ks.  We further processed the spectral data produced by \texttt{xselect} using the FTOOL \texttt{grppha}, which defined the binning and quality flags of the spectra.  We used the quality flag to further eliminate bad data.  Initially, we grouped the bins to ensure a minimum of 20 counts to fit the spectra using $\chi^2$ statistics.  However, insufficient bins were produced in the resulting spectrum.  Due to the small number of counts, we therefore used the ``C'' statistic \citep{Cash1979} for the spectral analysis.  The cumulative spectrum was grouped to have 5 counts per bin.
}

\subsection{Radio Observations and Analysis\label{sect:radio_obs}}

Radio observations were obtained using the Australia Telescope Compact Array \citep[ATCA;][]{Wilson2011}. 
Dedicated follow-up observations were made between 2017 November 17 and 2018 December 8 
(MJD 58,074 to 58,460, see Table \ref{table:atca}) with observations centered at 5.5 and 9.0 GHz, 
with 2 GHz bandwidths for both bands. The ATCA, which consists of six 22 m-diameter antennas, 
was in several different array configurations over this period, with the more compact arrays
somewhat more sensitive to the bright extended emission in the vicinity. Details of
the array configurations are given in Table \ref{table:atca}.

Observations were reduced following standard procedures in Miriad \citep{Sault1995}, 
with the flux density scale set by
observations of calibrators PKS 1934-638 and/or PKS 0823-500.
Initially, observations were made covering the nominal position of the 3FGL source, and the positions
of radio sources cataloged by \citet{Schinzel2017} and candidate X-ray sources.
Observations differed in length, hour-angle coverage, angular resolution, and sensitivity 
to extended radio emission in the vicinity, resulting in a heterogeneous data set.
\citet{Schinzel2017}, who observed with the ATCA in the relatively compact H168 and H214 array
configurations, detected four sources within the 3$\sigma$ error region around \src,
all of which were brighter than 15 mJy, and are resolved out on longer ATCA baselines.
\citet{Schinzel2017} also list a further 60 objects in the field detected 
outside the 3$\sigma$ region, which
appear to be associated with the HII region 
cataloged by \citet{Caswell1987} at (l,b) = (311.627, +0.27).
Observations of the candidate X-ray sources in more extended ATCA array configurations
revealed the fainter (\sqig2 mJy), variable radio counterpart.
The final observations of the series were conducted as targeted observations, 
with the phase calibrator PKS 1420-679 used throughout.


\subsection{Near-Infrared Observations and Analysis\label{sect:ir_obs}}

We observed the candidate near-IR counterpart to \src\ \citep[2MASS J14051441$-$6118282, allWISE J140514.40$-$611827.7;][]{Landi2017}
with FLAMINGOS-2 \citep{Eikenberry2004} 
on Gemini South on 2018 June 7 through Program ID GS-2018A-Q-412. We used the R3K grism and a 
4-pixel slit, giving a resolution of $R\sim1800$ across the $K$-band. The target was observed 
with 10 120-sec exposures in a standard ABBA nod, giving 1200 sec of exposure time total. 
We also observed the nearby A0V star HD 119942 as a telluric standard immediately before our science observations.

The data were reduced using standard near-IR procedures, with optimal extraction of the spectra from subtracted, flat-fielded nodded pairs. The final combined spectrum has a signal-to-noise of about 140 per resolution element in the continuum. After interpolating over the intrinsic Br$\gamma$ feature in the telluric standard, we corrected the \src\ spectrum for telluric absorption by scaling the telluric spectrum to minimize the residuals in the regions of strong telluric lines.

\section{Results\label{sect:results}}

\subsection{Gamma-ray Results\label{sect:gray_results}}

The power spectra of the light curves of all 3033 3FGL and 5523 FL8Y sources were examined
for indications of periodic modulation that could arise from previously unknown
binary systems.
For candidate new binaries our usual threshold for further investigation has been for a source
to have a peak power $\ge$ 18 $\times$ mean power level (FAP $<$ 5$\times$10$^{-4}$)
and for the period not to coincide with a known artifact. We also exclude
very long periods that are suggestive of arising from red-noise type behavior,
such as from active galactic nuclei.

\begin{figure}
\includegraphics[width=7.25cm,angle=0]{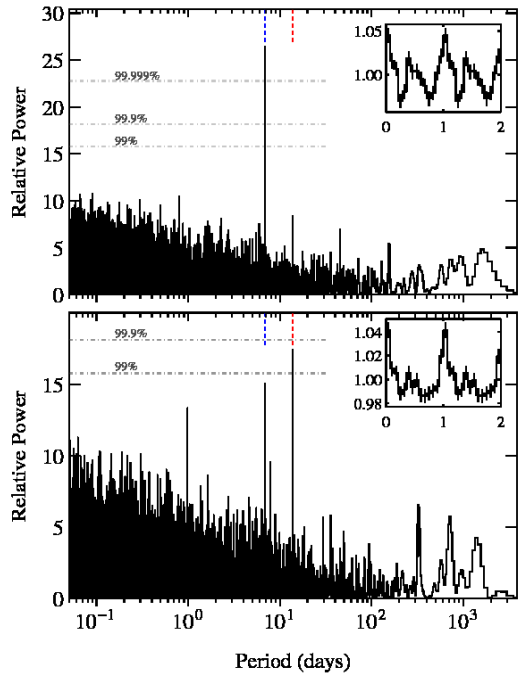}
\caption{Power spectra of LAT light curves of \src.
{\bfc The bottom panel shows the power spectrum of 
the probability-weighted LAT light curve (E $>$ 100 MeV) while
the upper panel shows the power spectrum from conventional aperture photometry without probability weighting
(E $>$ 200 MeV).
In both panels the vertical dashed lines indicate the harmonically
related peaks at 13.7135 $\pm$ 0.0019 (red) and 6.85675 $\pm$ 0.00096 (blue) days.
The peak at 1 day in the lower panel is an artifact commonly seen in LAT power spectra.
The dot-dashed horizontal gray lines show the white-noise significance levels.
The insets show the light curves folded on a 13.7135 day period.
For clarity, two cycles are shown, and the count rates have been normalized by
dividing by the mean rate.
}
}
\label{fig:double_power}
\end{figure}

From our power spectra, initially from 3FGL sources, and persisting into the FL8Y list
and the 4FGL catalog \citep{FermiLAT2019}, we noted
two peaks in the power spectra of {\bfa \three_src\ } and \flsrc\ which, although lower than our usual
threshold for further investigation, were consistent at a 1.7$\sigma$ level with being harmonics.
From the power spectrum of \src\ {\bfc (Fig. \ref{fig:double_power}, bottom panel)} the two peaks are at 6.8586 $\pm$ 0.0013 days
and 13.7235 $\pm$ 0.0046 days. The heights of the peaks are \sqig15.1 and \sqig17.4 compared to
the mean power level respectively,
with FAP values of 0.02 and 0.002, for 77436 trials. For a two value trial, i.e. performing
a test on whether there is modulation at either twice or half the period of stronger peak,
then the peak height of the 6.86 day period implies an FAP of 6$\times$10$^{-7}$.

Folding the LAT light curve on the 13.7 day period {\bfc (Fig. \ref{fig:double_power}, bottom panel inset)} shows a profile
with a single sharp peak, which occurs near MJD 56,498.7.
The presence of two peaks in the power spectrum can be
ascribed to the sharpness of this peak which deviates from a simple sine modulation.
Since the probability-weighted light curve can potentially suffer from photometric
distortions (Section \ref{sect:gray_obs}) we also created a light curve using conventional
aperture photometry without the weighting. For this we employed a one degree radius
aperture. From the power spectrum of this (Fig. \ref{fig:double_power}, top panel) we find instead
a single peak at 6.85675 $\pm$ 0.00096 days
and only a very small peak near 13.7 days.
The height of the peak for the power spectrum of the unweighted $>$ 200 MeV
light curve is 26.5 with an associated FAP of 3$\times$10$^{-7}$, allowing
only for the number of independent frequencies.
Through the remainder of this paper we use a period of 13.7135 $\pm$ 0.0019 days,
as the period implied by the harmonic in the power spectrum of the unweighted
light curve.
{\bfc
Period search techniques based on light-curve folding such as $\chi^2$ maximization \citep[e.g.][]{Leahy1983}
or phase-dispersion
minimization \citep{Stellingwerf1978}
can naturally also produce signals at ``sub-harmonics'' of the intrinsic modulation
period, i.e. multiples of the intrinsic modulation period. However, this effect
does not occur in direct Fourier-based analyses {\bfd which quantify
the sine-wave components of the modulation}. For this reason, we identify the
longer, \sqig13.7 day, period as the intrinsic modulation period of \src.
This is confirmed by the single peak exhibited in the X-ray and radio light curves
when folded on the longer period (Sections \ref{sect:xray_results} and \ref{sect:radio_results} respectively).
{\bfd The stronger peak near 13.7 days in the power spectrum of the conventional photometry
light curve compared to the probability-weighted light curve is because the modulation in the
conventional light curve is more nearly sinusoidal on this period. For the probability-weighted
light curve, the modulation profile is more sharply peaked (i.e. less sinusoidal), 
and thus the power is spread over more than one Fourier component, principally the fundamental 
and the first harmonic.
Such a change in profile shape between unweighted and weighted light curves is not necessarily
expected to be a general feature of modulation in HMGBs.
}
We note that if only the power spectrum of the conventional aperture photometry LAT light curve
was available (Fig. \ref{fig:double_power}, bottom panel) then the shorter harmonic period might have been incorrectly 
determined to be the period of the system.
}

The unweighted light curve folded on 13.7 days {\bfc (Fig. \ref{fig:double_power}, top panel)} 
now shows, in addition to the primary peak at phase 0, an additional
prominent peak near phase 0.5. 
We then examined the modulation as a function of energy. 
We find that below 200 MeV no modulation is detected. Above
200 MeV we find that there is a difference in the two peaks,
where one is more prominent at lower energies, while the other
becomes more prominent at higher energies, and coincides with
the phasing of the single peak in the probability-weighted light curve.
Thus, we ascribe the difference between the power spectra for the
weighted and unweighted light curves as an energy-dependent effect,
where the high-energy photons are effectively over-weighted because
of the smaller point-spread function.

\begin{figure}
\includegraphics[width=7.25cm,angle=0]{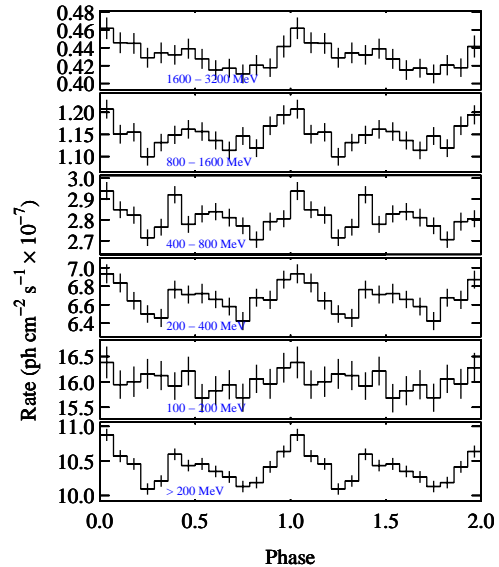}
\caption{Conventional aperture photometry of LAT observations of \src\
folded on the 13.7 day period separated by energy.
The folded light curves are not background subtracted.
Phase zero corresponds to MJD 56,498.7.
}
\label{fig:multi_e_fold}
\end{figure}

{\bfc
To investigate the stability of the period, we divided the conventional
aperture photometry light curve into
three equal-length sections and calculated the power spectrum of each
section separately. The periods of the orbital harmonic derived in this way (6.855 $\pm$ 0.006,
6.860 $\pm$ 0.004, and 6.853  $\pm$ 0.005 days) are all consistent with each other,
and also the more precise period derived from the entire light curve.
}

{\bfc
Neither the power spectrum of the probability-weighted light
curve nor that of the unweighted light curve (Fig.\ref{fig:double_power})
show any evidence of additional periodicities, or long-period/low-frequency variability,
beyond the two peaks related to the \sqig 13.7 day period, and an artifact at one day
that is often seen in our power spectra of LAT light curves.
}

{\bfc The spectral model derived for \src\ in the 4FGL catalog is a log normal function (\texttt{LogParabola}), 
as it was in the 3FGL catalog. i.e.
\begin{equation}
\frac{dN}{dE} = K \left(\frac{E}{E_0}\right)^{-\alpha-\beta\,log_e(E/E_0)}
\end{equation}
This model is used in the LAT catalogs for all sources with significantly
curved spectra.  
Additional information on the 3FGL and 4FGL catalog results, including
plots of spectral fits are available from the Fermi Science Support Center
\footnote{https://fermi.gsfc.nasa.gov/ssc/data/access/lat/4yr\_catalog/}
\footnote{https://fermi.gsfc.nasa.gov/ssc/data/access/lat/8yr\_catalog/}.
In Table \ref{table:hmgb_spectra} we show the spectral parameters in the 4FGL catalog of
\src\ together with the other persistent HMGBs. In all cases the model employed
in the catalog was \texttt{LogParabola}, and
we note that the spectral parameters of all sources are broadly comparable.
}

\begin{figure}
\includegraphics[width=7.25cm,angle=270]{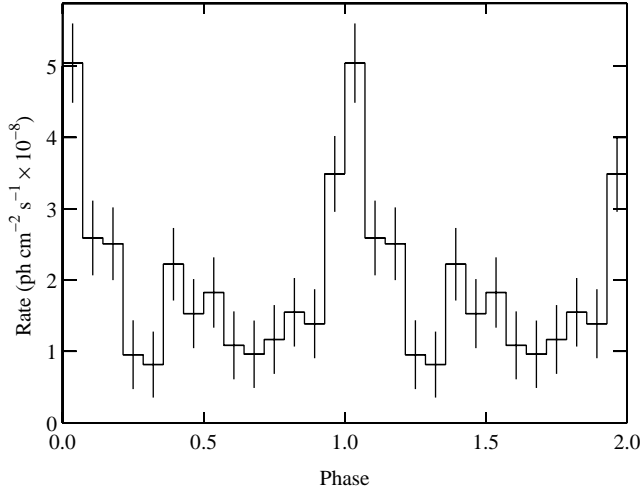}
\caption{Gamma-ray flux of \src\ obtained from a phase-resolved likelihood
analysis of the LAT data for energies between 200 MeV to 500 GeV. 
Spectral parameters were frozen to the values from the 4FGL catalog.
}
\label{fig:phase_likelihood}
\end{figure}

We made a phase-resolved maximum likelihood analysis to also examine the
modulation on 13.7 days. For this we generate {\bfc a model} based on the 4FGL
source catalog, but held all parameters fixed, apart from the flux of \src.
We employed a 10 degree radius ``region of interest''.
The {\bfa phase-resolved fluxes} derived from this likelihood analysis are plotted in Fig. \ref{fig:phase_likelihood}.
We find that the flux increase near phase zero is clearly detected, but
only a modest increase near a phase of 0.5 is observed. 
The overall folded light curve also appears somewhat noisier with a lower
amplitude than the folded aperture photometry light curve.
We investigated phase-resolved likelihood analysis for energy ranges between
200 to 1000 {\bfb MeV}, and 1000 to 500,000 {\bfb MeV}. However, the fits to the lower energy
range resulted in low Test Score (TS) numbers and so could not be used to
investigate the secondary peak. 
We also made fits with the spectral parameters of \flsrc\ left free,
but these were found to result in spectral parameters which varied
in an implausible way between phase bins.

While the likelihood analysis does not strongly show the double-peaked
orbital profile shown by the conventional aperture photometry, we note that
there is no known artifact that would produce double-peaked modulation in the aperture
photometry.
In addition, the aperture photometry has the benefit that
it is model independent. For these reasons, along with the modest secondary peak
that is seen in the likelihood analysis, we believe that the double-peaked
structure more strongly found in the aperture photometry is indeed showing the intrinsic
behavior of the source, while the secondary apparently softer peak is more difficult to fit
due to the higher background at lower energies.

\subsection{X-ray Results\label{sect:xray_results}}
\subsubsection{X-ray Flux Variations}

\begin{figure}
\includegraphics[width=7.25cm,angle=270]{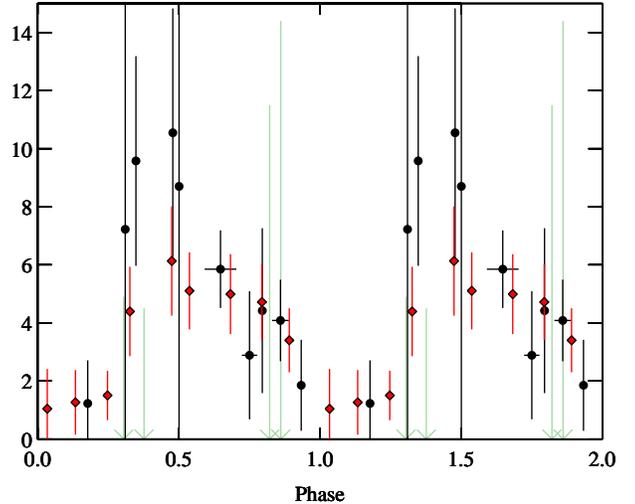}
\caption{\Swift\ XRT flux of the counterpart of \src\
folded on the 13.7 day period. The archival observations are shown
as black circles, while our Target of Opportunity observations are shown as red
diamonds. 
{\bfa
The green arrows are upper limits (3 $\sigma$) from the archival observations. }
}
\label{fig:xray_fold}
\end{figure}

{\bfa
The XRT TOO observations of \src\ cover more than two orbital periods.
}
The XRT light curve folded on the 13.7 day period is shown in Fig. \ref{fig:xray_fold}.
From this, a strong, approximately sinusoidal, modulation on the 13.7 day
gamma-ray period can be seen.
However, we note that X-ray minimum occurs near the phase of primary gamma-ray maximum.
Although the archival observations have shorter durations, and thus larger error bars,
than our TOO observations, they also indicate modulation on the 13.7 day period.

{\bfa
To investigate the energy dependent phase shift between the \textsl{Fermi} gamma-ray and 
\textsl{Swift} X-ray light curves, we calculated the maximum value of the cross correlation function 
between the two folded light curves
after applying phase shifts to the X-ray light curve. For this we only used XRT data from the TOO observations.
Due to the sparse coverage of the X-ray light curve, we did not bin this but instead
linearly interpolated to cover observation gaps.
From this we find the X-ray light curve leads the LAT light curve
by a phase of 0.586.
We also used a simple sine wave fit to the entire XRT light curve to determine maximum of the X-ray light curve 
and find this occurs at a phase
of 0.59 $\pm$ 0.03. This is thus consistent with the results of the cross-correlation analysis.
However, this result should be treated with caution since we only have good coverage of two cycles,
the individual flux measurements have relatively large errors, and the shape of the modulation is hence not yet well
defined.
}

{\bfa
\subsubsection{X-ray Spectrum}

The cumulative and orbital-peak spectra of \src\ were analyzed using \texttt{XSPEC v12.9.0k}.  We made use of the \texttt{XSPEC} convolution model \texttt{cflux} to calculate the fluxes and associated errors of \src.  To fit the cumulative {XRT} spectrum, we used several models that are used to describe systems that host a neutron star: a power law (\texttt{power}), a power law with a high-energy cutoff (\texttt{highecut} in XSPEC), and a cutoff power law (\texttt{cutoffpl} in XSPEC).  All models were modified by an absorber that fully covers the source using appropriate 
cross sections \citep{Verner1996} and abundances \citep{Wilms2000}.

We initially allowed the neutral hydrogen column density and photon index to be free parameters and performed spectral fits on the cumulative spectrum. Due to the short exposure times of the archival observations (see Table~\ref{table:xrt}), we chose to include only the TOO observations that were performed between 2018 May 16 and 2018 June 17 (MJD\,58,254 and MJD\,58,286).  The model that provides the best fit (C statistic of 10.19 for 18 degrees of freedom) to the data is a power law.  While a good fit does not require a high-energy cutoff, which is 
typically found in accreting pulsars \citep{Coburn2002}, our spectra are limited to energies below 10\,keV.  We note that while high-energy cutoffs are typically seen at higher energies, none was found in LS 5039 \citep{Takahashi2009} or 
1FGL J1018.6-5856 \citep{An2015} up to 40\,keV.  
We find the neutral hydrogen column density for the fully covered absorption and the photon index to be 1.1$^{+0.9}_{-0.8}$$\times$10$^{23}$\,cm$^{-2}$ and 2\,$\pm$\,1, respectively.  We also attempted fitting the cumulative spectrum of the TOO observations along with the archival ones and found larger uncertainties in the model parameters.  
The cumulative XRT spectrum, and best-fitting power-law model are shown in Fig. \ref{fig:xrt_spectrum}.

\begin{figure}
\includegraphics[width=7.25cm,angle=0]{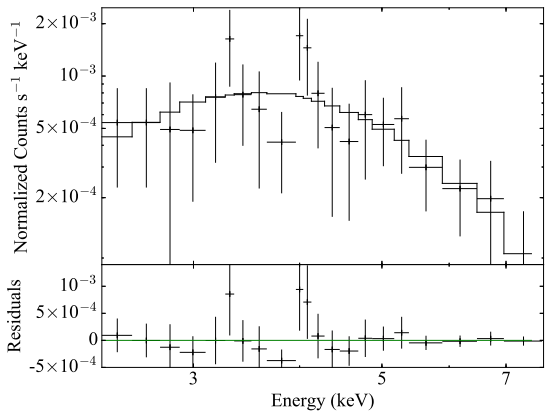}
\caption{
{\bfa
Cumulative \textsl{Swift} XRT spectrum of \src\ (= \cxou).  The best fit power-law model is shown as
a histogram.  
}
}
\label{fig:xrt_spectrum}
\end{figure}

Since neither the neutral hydrogen column density for the fully covered absorber nor the photon index could be accurately constrained, 
we converted the mean derived value of optical reddening $E(B-V)$ of 10.2 (Section \ref{sect:ir_results}) into the predicted neutral hydrogen column density 
and chose to fit the cumulative spectrum with the $N_{\rm H}$ frozen to it.  From Equation 1 in \citet{Guver2009}, 
we calculate the neutral hydrogen column density along the line of sight to be 6.9$\times$10$^{22}$\,cm$^{-2}$.  
Using the \texttt{power} model, we find the photon index to be 1.5\,$\pm$\,0.4
and the unabsorbed X-ray flux in the 2--8\,keV band to be 8\,$\pm$\,1$\times$10$^{-13}$\,\ergcm2s. 
{\bfb For comparison, \citet{Takahashi2009} found a power law index for LS 5039 that varied between 1.45 - 1.61,
for 1FGL J1018.6 \citet{An2015} found indices between \sqig 1.4 - 1.7, and for LMC P3 indices between \sqig 1.3 to 1.6 have
been reported \citep{Bamba2006,Seward2012,Corbet2016}.}
It is important to note that while this {\bfb model}
provides a good fit to the XRT spectrum (C statistic of 10.39 for 19 degrees of freedom), {\bfc this interpretation 
should be treated with caution}
since we assumed that the fully covered $N_{\rm H}$ is entirely interstellar in origin.  While this is the simplest 
interpretation of the data, we cannot exclude the possibility that the fully covered $N_{\rm H}$ is due to a combination 
of intrinsic and interstellar absorbers. 
{\bfc
However, for LS 5039 and 1FGL J1018.6-5856 the measured $N_{\rm H}$ values are consistent with being
only due to interstellar absorption \citet{An2015,Takahashi2009}.
}

}

\subsection{Radio Results\label{sect:radio_results}}

A point {\bfa radio} source was found {\bfa with ATCA} in the LAT error region of the gamma-ray source
which {\bfa coincide}s with the position of \cxou\ and also the
proposed near-IR counterpart {\bfa at
R.A. = }14$^{\rm h}$05$^{\rm m}$14$^{\rm s}_.$42 $\pm$ 0.02",
{\bfa decl.} = $-$61$\degr$18$\arcmin$28$\arcsec_.$33 $\pm$ 0.03$\arcsec$ (J2000).
The offset of this position from the 3FGL, FL8Y, and 4FGL positions and corresponding
95\% semi-major error radii are 1.43 (1.88), 1.23 (1.31), and 1.50 (1.25)\arcmin\ respectively. 

\begin{figure}
\includegraphics[width=7.25cm,angle=270]{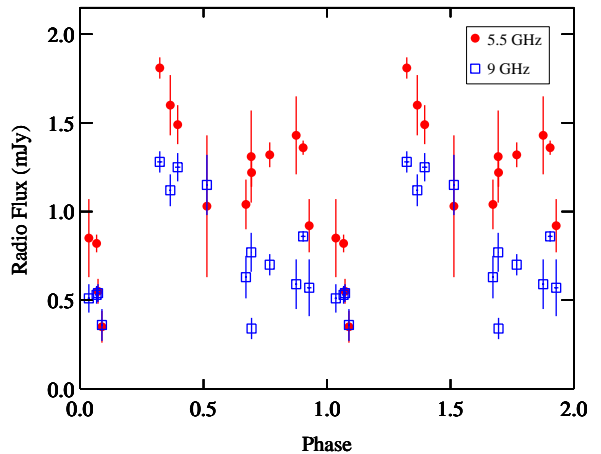}
\caption{ATCA observations of the counterpart of \src\
folded on the 13.7 day period.}
\label{fig:radio_fold}
\end{figure}

The radio flux densities folded on the 13.7 day period (Fig. \ref{fig:radio_fold}) 
show modulation of the emission on the gamma-ray period.
There is a gap in the coverage between phases \sqig0.1 and \sqig0.3,
which hampers our determination of the phase of maximum flux, but it must be
either within this gap or near phase \sqig0.3. We do have coverage
at phase 0, the maximum of the gamma-ray flux, and find a minimum radio
flux.
{\bfa
Thus, similar to the X-ray flux, the radio flux is offset from the primary
gamma-ray maximum. However, there is an indication that the radio flux maximum
may occur somewhat before the X-ray maximum.
}
We note a low intensity 9 GHz measurement at a phase of 0.7,
while the flux at 5.5 GHz is not exceptionally low.
The 9 GHz measurement appears to be reliable and not due
to an instrumental effect.

\subsection{Near-Infrared Results: Spectral Classification and Distance\label{sect:ir_results}}

\begin{figure}
\includegraphics[width=7.25cm,angle=0]{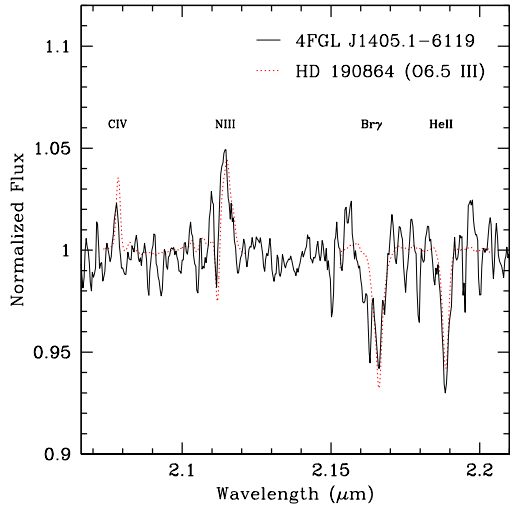}
\caption{Near-infrared spectrum of the counterpart of \src\ (2MASS J14051441$-$6118282, allWISE J140514.40$-$611827.7) 
obtained with FLAMINGOS-2 on Gemini South.
Overplotted as a red dashed line is the spectrum of the spectral comparison star HD 190864.
}
\label{fig:ir_spec}
\end{figure}

The near-infrared spectrum obtained with FLAMINGOS-2 is shown
in Fig. \ref{fig:ir_spec}.
To classify \src, we use the near-IR library of early-type stars from \citet{Hanson2005}. These spectra were obtained at higher resolution than our spectrum ($R\sim12000$ rather than $\sim1800$), so we smooth them to match our resolution before comparison.

The most prominent line in the spectrum is \ion{N}{3} in emission (rest frame 2.1155 $\mu$m). and we also observe weak \ion{C}{4} emission at 2.078 $\mu$m. Absorption lines clearly present are Br$\gamma$ (H) at 2.1661 $\mu$m and \ion{He}{2} at 2.188 $\mu$m, and a weaker \ion{He}{1} absorption line is present at 2.113 $\mu$m, nearly blended with the \ion{N}{3} emission line.

In terms of the lines present and absent and their relative strengths, the best match  from the \citet{Hanson2005} catalog is HD 190864, which is classified as an O6.5 III star. The object spectrum overplotted with this star is in Figure \ref{fig:ir_spec}. By comparison, an O5 III star in the catalog, HD 15558, has much weaker \ion{He}{2} than \src. Similarly, the O5/O5.5 supergiants in the catalog have weaker \ion{He}{2} and stronger \ion{C}{4} emission than \src. Considering O dwarfs, the absorption lines generally appear to be broader than observed in \src, consistent with their higher rotational velocities and/or gravities.
  
We conclude that, on the basis of our spectrum, the best classification of \src\ is as an  O6.5 III star, but emphasize that this classification should be revisited with a higher signal-to-noise spectrum in the future.
  
By cross-correlation between the standards and our spectrum in the region of the \ion{He}{2} line, we find a barycentric radial velocity 
{\bfa of $207\pm16$ km s$^{-1}$}. We also note that the emission lines appear to lie at a different velocity, so the systemic velocity of this star is not necessarily well-determined.

In order to estimate the distance to \src\ we used the tabulated absolute magnitudes from
\citet{Martins2006} for an O6.5 III star (i.e. M$_J$ = -5.03, M$_H$ = -4.92, and M$_K$ = -4.82), 
the 2MASS photometric measurements \citep[H = 14.369 $\pm$ 0.068, K = 12.769 $\pm$ 0.044][]{Landi2017},
and the reddening laws determined by \citet[][R$_V$=3.1]{Rieke1985} and also \citet{Schlafly2011}.
The infrared counterpart is also present in the Vista Variables in
the Via Lactea (VVV) catalog \citep{Minniti2010} with measured values of 
J = 17.270 $\pm$ 0.028 H = 14.524 $\pm$ 0.005 K = 12.854 $\pm$ 0.003.
The distances derived from these photometric measurements and reddening
curves are given in Table \ref{table:distance}.
The mean of these distances is 7.7 kpc, with a standard deviation
of 1 kpc. For the remainder of this paper we use 7.7 kpc for
estimates of source luminosity, but caution that there is considerable
uncertainty on this.
The mean derived value of E(B-V) is 10.2 with a standard deviation of 1.5.
We note that the reddening curve of \citet{Rieke1985} gives somewhat higher distances
and lower reddening than using the reddening curve from \citet{Schlafly2011}.
{\bfa
In addition, the 2MASS and VVV measurements are formally inconsistent. If this
is not due to, for example, different photometric bandpasses, it might indicate
variability in the near-IR.
}

%


\section{Discussion\label{sect:discuss}}

\subsection{Properties and Nature of \src\label{sect:discuss_properties}}

The periodic modulation on 13.7 days found at gamma-ray, X-ray, and radio
wavelengths, together with the identification of a stellar counterpart,
clearly show that \src\ is a high-mass gamma-ray binary, and the 13.7 day
period is expected to be the orbital period of the system.
\src\ is the third gamma-ray binary, after 1FGL J1018.6-5856 and LMC P3, 
to be found from the initial discovery of periodic modulation of the LAT
light curve. In all three cases, the primary star in the system is
an early O star, rather than a Be star. 
{\bfa
The O6.5 III classification for \src\ indicates a mass for the primary star
\sqig 25 - 35 M$_\sun$ \citep{Mahy2015}.
}

For a distance of \sqig 7.7 kpc, the implied maximum gamma-ray luminosity would be comparable
to, and possibly higher than, that of LS 5039, while approximately half that of 1FGL J1018.6-5856 and
a tenth that of LMC P3. 
The unabsorbed mean X-ray luminosity of \src\ is \sqig5.6$\times$10$^{33}$\ergs\,($d$/7.7 kpc)$^2$.
For LMC P3 the unabsorbed luminosity measured with the \Swift\ XRT (2.0 - 7.5 keV)
was also considerably larger at \sqig9.6$\times$10$^{34}$\ergs\,($d$/50 kpc)$^2$ \citep{Corbet2016}.
For 1FGL J1018.6-5856 there are flares that reach unabsorbed luminosities (0.5 - 10 keV) 
of \sqig10$^{34}$\ergs\,($d$/5.4 kpc)$^2$,
while between these the luminosity can decline to \sqig2.6$\times$10$^{33}$\ergs\,($d$/5.4 kpc)$^2$ \citep{An2013}
and so on average it is more similar to \src. For LS 5039, the X-ray luminosity (1 - 10 keV) is also comparable at
\sqig6$\times$10$^{33}$\ergs\,($d$/2.5 kpc)$^2$ 
{\bfc
\citep{Bosch2007,Takahashi2009,Rea2011}.
}

Because of the similarities with the other systems, we also hypothesize that gamma-ray
emission from \src\ is driven by the interaction between the wind from a rapidly rotating
neutron star and the wind from the O star companion.
Orbital modulation of observed flux in HMGBs can be driven by orbital phase-dependent
changes in both viewing angle and varying source distance for systems with significant
eccentricity. The overall pattern of variability in \src\ is similar to that seen
in LMC P3 where the X-ray and radio modulations are approximately in phase with
each other, but close to 0.5 out of phase from the primary gamma-ray peak.
{\bfa However, the energy-dependent gamma-ray modulation seen in \src\ with a secondary
softer peak approximately 0.5 out of phase from the primary peak is unlike
LMC P3. For \J1018\ the X-ray emission has a broad component that is out of phase
from the overall gamma-ray modulation, although there is an additional X-ray component
that exhibits sharp ``flares'' that is in phase with the gamma-ray modulation \citep{FermiLAT2012,An2015}.
\citet{An2017} have also reported for \J1018\ that the gamma-ray light curve below 200 MeV includes
a component that is modulated with a similar phasing to that of the broad X-ray component.
}
Anti-phasing between the modulation of the gamma-ray and X-ray emission in the O-star {\bfc HMGBs} may be explained
if the gamma-ray variability is primarily due to inverse Compton scattering, which
gives greater flux near superior conjunction, and X-ray modulation is due
to Doppler boosting, which will give higher observed flux near inferior conjunction \citep[e.g.][]{Dubus2015b}.

To investigate the driving mechanisms behind these modulations in detail it will
be important to determine a radial velocity curve for the system.
In addition to the orbital variability of the radio flux, there also appears to
be shorter-term variability which does not occur simultaneously at both
frequencies.
The origin of this radio variability is unclear, but it {\bfc may }be similar to what has
been seen in LMC P3 \citep{Corbet2016}.

We note that \src\ is located \sqig40\arcmin\ from
the
previously suggested SNR counterpart G311.5-00.3 {\bfa \citep{Saz2016}}, which has
a 5\arcmin\ radio diameter \citep{Green2014}, and so the two sources must be distinct.
Thus, for \src\ we cannot directly estimate the age of the system from an
associated SNR.
Similarly, for 1FGL J1018.6-5856 \citet{Marcote2018} also find that their motion
and distance measurements are not consistent with a previously
proposed possible SNR association. This is unlike LMC P3 for which
there does appear to be an association with an SNR \citep{Seward2012,Corbet2016}.

{\bfc
The Be star HMGB \lsi\ have shown, in addition to its \sqig 26.5 day orbital period,
a superorbital period near 1667 days that was originally detected from
radio observations \citep[e.g.][and references therein]{Gregory2002}. 
This superorbital modulation has also been seen in the LAT light curve of
this source \citep[e.g.][]{Hadasch2012,Xing2017,Jaron2018}.
Such superorbital variability has not yet been shown to be a general
property of HMGBs and, as noted in Section \ref{sect:gray_results},
we do not see such modulation in \src. However, since superorbital
modulation has been reported in a number of high-mass X-ray binaries accreting from both Be star envelopes and
the winds from OB star companions 
\citep{Raj2011,Corbet2013,Corbet2018},
it is possible that superorbital modulation may still be found in other HMGBs beyond
\lsi.
}

\subsection{Distance Limits on Detectability of Galactic Gamma-ray Binaries\label{sect:discuss_population}}

The Galactic population of gamma-ray binaries has been discussed by \citet{Dubus2017} and
these authors found very large uncertainties on the possible population as $101_{-52}^{+89}$.
We previously suggested that the discovery of a binary in the LMC but the lack of
detection, at that time, of additional Galactic systems might suggest we had
discovered all detectable sources. Here we investigate this in more detail
by estimating the approximate maximum distance at which we could
detect the known binaries from their periodic variability in LAT light curves.

\begin{figure}
\includegraphics[width=7.25cm,angle=270]{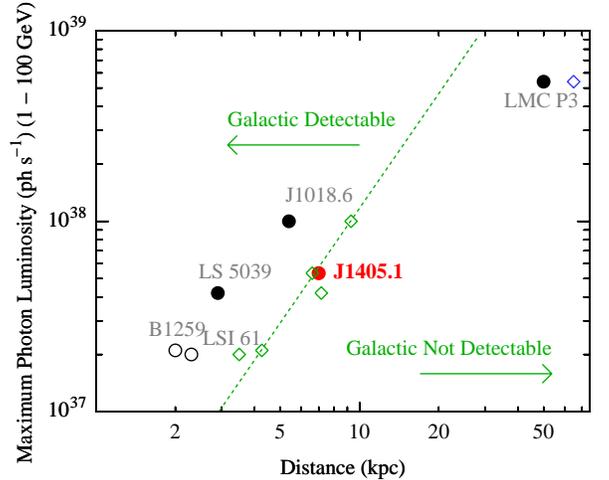}
\caption{Maximum gamma-ray photon luminosity plotted against
distance for HMGBs that show detectable periodic
modulation in their LAT light curves. Sources with
Be star primaries are plotted as open circles, sources with
O star primaries as filled circles.
The green
diamonds show the approximate maximum distance
at which modulation could be detected in each source, see text for details.
The green line is a fit to the maximum detectable distances
with a slope fixed equivalent to {\bfb distance}$^{-2}$.
The use of a different distance for an individual source would cause
its location in this diagram to move parallel to the green
dashed line.
}
\label{fig:l_vs_d}
\end{figure}

In Fig. \ref{fig:l_vs_d}, using Table 3 of \citet{Dubus2017} together with
the parameters of \src, we plot photon
luminosity of those sources with detectable modulation against distance.
We note that there is a general trend for the most distant known sources to
be the most luminous. This is suggestive of a flux detection threshold.

As a detection threshold in our power spectra, we consider that a peak
should have a minimum height of 20 times the mean power level to be
detectable. We then determined the actual heights of the orbital
peaks in the {\bfa FL8Y} LAT light curves using light curves covering from 
2008 August 5 and 2018 July 5 (MJD 54,683
to 58,304). The ratio of the maximum detection distance to the actual
distance was then taken as (peak height/20)$^{1/4}$. One square root
to convert from power to amplitude, and one square root to allow
for inverse square law decrease in flux. We then multiplied the
source distances by the resulting factor, and these are plotted
as green diamonds in Fig. \ref{fig:l_vs_d}. The green line is
a straight line fit (in log-log space) with a slope fixed to 
be equivalent to r$^{-2}$. The space to the left of
this line is assumed to be detectable, while the space to the right
would not be detectable. Note that changed distance measurements
would move source locations along lines parallel to the green
line, and thus the location of the green line itself would not change.
In addition, due to the fourth root dependence, the location of the line
does not strongly depend on the choice of detection threshold relative power level. 

Although our binary detection limit is derived in a simple way, and does
not account for different modulation patterns
or differing backgrounds, it does give generally consistent
results for the Galactic binaries. We note, however, that LMC P3
does lie inside the region where it would nominally not be
detectable. Thus, the detectability of LMC P3 may be due to
the considerably lower background {\bfa due to it} being located away from the
Galactic plane. The Be star system \hessj, which is
at most weakly detected with the LAT, and so does not exhibit
periodic modulation, is at a distance of 0.8 kpc and a luminosity
of 2.9$\times$10$^{35}$ ph\,s$^{-1}$ \citep{Dubus2017} 
and hence is located to the right of the nominal detection threshold line.

With extended LAT observations, the detection line will slowly move
to the right. We find {\bfa empirically} that the strengths of the peaks of the known periods in the
power spectra relative to the mean power levels increase approximately linearly with time
{\bfa (i.e. the mean amplitude noise decreases as the square root of time)}.
Thus, the maximum detection distance would increase as $\propto t^{-4}$.
The maximum detection volume, for sources near the plane, would increase
as $\propto t^{-2}$. Convolved with this, for the discovery rate of
new binaries from their modulation, is the poorly constrained
luminosity function for gamma-ray binaries. We note that the
Be star systems are relatively nearby, and lower luminosity,
perhaps suggestive that there remains a more extensive
population of Be stars remaining to be discovered.

Our searches for binaries based on detection of periodic
modulation do require us to distinguish between periodic
and non-periodic variability. If we require $\gtrsim$10 cycles
to enable clear discrimination between a coherent signal
and red noise, then periods $\lesssim$1 year would be required.
Thus, discovery of Be star binaries with periods such as
those of PSR B1259-63 and PSR J2032+4127 would be difficult or impossible.
However, Be X-ray binaries generally exhibit binary periods
between \sqig20 to \sqig 400 days \citep[e.g.][and references therein]{Corbet2017}.
Brief flaring behavior such as exhibited by PSR B1259-63 \citep[e.g.][]{Johnson2018} 
would also be more readily distinguished from red noise behavior with fewer cycles.

From the modest number of gamma-ray binaries known so far, there are two factors
that may facilitate the discovery of O-star systems from the detection
of modulated gamma-ray emission. The orbital periods of
the O-star systems are shorter than those of the known Be systems.
In addition, the Be star systems typically have lower luminosities
and so the O-star systems are visible at greater distances.

{\bfa
Our detection threshold distance for sources with similar luminosities and
gamma-ray modulation properties to LS 5039 and \src\ raises the possibility that 
the HMGB population estimate of \citet{Dubus2017} may have been an underestimate.
\citet{Dubus2017} estimated that \sqig80\% of LS 5039-like systems in the Galaxy
would be detected. However, discovery from gamma-ray modulation alone within the
Galactic plane appears currently restricted to \sqig 7 kpc. Thus, multi-wavelength
observations of unidentified gamma-ray sources to aid variability searches will continue to be important.
For example, the identification of an OB counterpart to an X-ray source {\bfa within} a LAT error ellipse
could also facilitate the discovery of a gamma-ray binary.
}

\section{Conclusion\label{sect:conclude}}

We have identified \src\ as a gamma-ray binary with gamma-ray flux
modulated on a period near 13.7 days. This periodicity is also
seen at X-ray and radio wavelengths. 
As with 1FGL J1018.6-5856 and LMC P3, which were also
detected from modulated gamma-ray emission with the LAT, \src\ contains
an O, rather than a Be, star primary.
The system is heavily obscured in the optical. Future infrared radial
velocity measurements would be valuable for determining the system
geometry and establishing the system orientation at maximum flux in
the different wavebands. A determination of the system eccentricity
would also be {\bfa important}.
It is suspected that \src\ contains a rapidly rotating neutron star,
although we do not yet have a direct detection of this.
The Galactic population of gamma-ray binaries is unclear, but there
may remain a lower gamma-ray luminosity population remaining to be
discovered, particularly since for the known systems, the Be
star systems have lower luminosities.

\acknowledgements

{\bfc We thank an anonymous referee for useful comments.}
This work was partially supported by NASA \Fermi\ grant 
NNX15AU83G.
The Australia Telescope Compact Array is part of the Australia Telescope National Facility which is 
funded by the Australian Government for operation as a National Facility managed 
by CSIRO. 
J. Strader acknowledges support from a Packard Fellowship. This work was partially supported by NASA grant 80NSSC17K0507 and NSF grant AST-1714825. This paper was partially based on observations obtained at the Gemini Observatory, which is operated by the Association of Universities for Research in Astronomy, Inc., under a cooperative agreement with the NSF on behalf of the Gemini partnership: the National Science Foundation (United States), the National Research Council (Canada), CONICYT (Chile), Ministerio de Ciencia, Tecnolog\'{i}a e Innovaci\'{o}n Productiva (Argentina), and Minist\'{e}rio da Ci\^{e}ncia, Tecnologia e Inova\c{c}\~{a}o (Brazil).
We thank the Swift team for undertaking observations.
The \textit{Fermi} LAT Collaboration acknowledges generous ongoing support
from a number of agencies and institutes that have supported both the
development and the operation of the LAT as well as scientific data analysis.
These include the National Aeronautics and Space Administration and the
Department of Energy in the United States, the Commissariat \`a l'Energie Atomique
and the Centre National de la Recherche Scientifique / Institut National de Physique
Nucl\'eaire et de Physique des Particules in France, the Agenzia Spaziale Italiana
and the Istituto Nazionale di Fisica Nucleare in Italy, the Ministry of Education,
Culture, Sports, Science and Technology (MEXT), High Energy Accelerator Research
Organization (KEK) and Japan Aerospace Exploration Agency (JAXA) in Japan, and
the K.~A.~Wallenberg Foundation, the Swedish Research Council and the
Swedish National Space Board in Sweden. Additional support for science analysis during the
operations phase is gratefully acknowledged from the Istituto Nazionale di Astrofisica in
Italy and the Centre National d'\'Etudes Spatiales in France.

\clearpage

\begin{deluxetable}{ccccccc}


\tablecolumns{7}
\tabletypesize{\small}
\tablewidth{0pc}
\tablecaption{Swift XRT Observation Log of CXOU J053600.0-673507 {\bfa (\src)}}
\tablehead{
\colhead{ObsID} & \colhead{Start Time (UT)} & \colhead{End Time (UT)} & \colhead{Phase$^a$} & \colhead{Exposure$^b$} & \colhead{Count Rate$^c$} & \colhead{Flux$^d$}}
\startdata
00041805001 & 2011-10-04 02:55:00 & 2011-10-04 23:05:29 & 0.831--0.890 & 3.1 & 4$_{-1}^{+2}$ & {\mybf 9$^{+4}_{-3}$} \\
00041805002 & 2011-10-08 20:56:00 & 2011-10-08 21:48:29 & 0.177--0.178 & 1.0 & 1$_{-1}^{+2}$ & {\mybf 3$^{+5}_{-2}$} \\
00042313001 & 2011-10-10 14:52:59 & 2011-10-10 15:44:59 & 0.304--0.305 & 0.5 & $<$4.9 & {\mybf $<$11} \\
00041805003 & 2011-11-01 22:55:00 & 2011-11-01 23:49:51 & 0.933--0.934 & 1.3 & 2$_{-1}^{+2}$ & {\mybf 4$^{+4}_{-3}$} \\
00042320001 & 2012-03-10 10:58:00 & 2012-03-10 11:49:44 & 0.376--0.377 & 0.6 & $<$4.5 & {\mybf $<$10} \\
00041805005 & 2012-09-21 09:30:59 & 2012-09-22 23:05:25 & 0.592--0.704 & 4.6 & 5\,$\pm$\,1 & {\mybf 13\,$\pm$\,3} \\
00084747001 & 2015-02-19 18:29:00 & 2015-02-19 18:39:00 & 0.861--0.861 & 0.6 & $<$14.4 & {\mybf $<$33} \\
00084747003 & 2016-12-04 01:34:00 & 2016-12-04 01:39:00 & 0.500--0.500 & 0.3 & 8$^{+14}_{-7}$ & {\mybf 20$^{+33}_{-16}$}\\
00084747004 & 2016-12-07 02:53:00 & 2016-12-07 20:41:00 & 0.722--0.777 & 1.1 & 3$^{+3}_{-2}$ & {\mybf 7$^{+7}_{-4}$}\\
00084747005 & 2016-12-15 03:54:00 & 2016-12-15 03:59:00 & 0.309--0.309 & 0.2 & 8$^{+13}_{-6}$ & {\mybf 18$^{+29}_{-14}$} \\
00084747006 & 2017-09-22 02:07:00 & 2017-09-22 02:25:00 & 0.794--0.795 & 1.0 & 4$^{+3}_{-2}$ & {\mybf 10$^{+8}_{-5}$} \\
00084747007 & 2017-12-07 05:30:00 & 2017-12-07 05:47:00 & 0.346--0.347 & 0.9 & 10$^{+4}_{-3}$ & {\mybf 22$^{+10}_{-8}$} \\
00084747008 & 2017-12-09 00:35:00 & 2017-12-09 00:48:00 & 0.477--0.478 & 0.7 & 10$^{+5}_{-4}$ & {\mybf 24$^{+10}_{-9}$} \\
00084747009 & 2017-12-13 17:50:00 & 2017-12-13 17:55:00 & 0.821--0.822 & 0.2 & $<$11.5 & {\mybf $<$27} \\
00084747010 & 2018-05-16 10:02:00 & 2018-05-16 13:35:00 & 0.027--0.038 & 3.9 & 1.0$^{+1.8}_{-0.9}$ & {\mybf 2$^{+4}_{-2}$} \\
00084747011 & 2018-05-19 08:09:00 & 2018-05-19 11:48:00 & 0.241--0.251 & 3.9 & 1.5$^{+1.0}_{-0.7}$ & {\mybf 4\,$\pm$\,2} \\
00084747012 & 2018-05-22 11:01:00 & 2018-05-22 14:46:00 & 0.468--0.479 & 3.9 & 6\,$\pm$\,2 & {\mybf 14$^{+5}_{-4}$} \\ 
00084747013 & 2018-05-25 07:43:00 & 2018-05-25 11:05:00 & 0.677--0.687 & 3.7 & 5\,$\pm$\,1 & {\mybf 12\,$\pm$\,3} \\ 
00084747014 & 2018-05-28 04:14:00 & 2018-05-28 07:36:00 & 0.885--0.895 & 3.8 & 3\,$\pm$\,1 & {\mybf 8$^{+3}_{-2}$} \\ 
00084747015 & 2018-05-31 08:51:00 & 2018-05-31 18:32:00 & 0.118--0.147 & 3.2 & 1.2$^{+1.4}_{-0.9}$ & {\mybf 3$^{+3}_{-2}$} \\ 
00084747016 & 2018-06-03 03:30:00 & 2018-06-03 06:53:00 & 0.320--0.330 & 3.9 & 4$^{+2}_{-1}$ & {\mybf 10$^{+4}_{-3}$} \\ 
00084747017 & 2018-06-06 00:21:00 & 2018-06-06 05:09:00 & 0.529--0.544 & 4.0 & 5\,$\pm$\,1 & {\mybf 12\,$\pm$\,3} \\ 
00084747018 & 2018-06-09 09:29:00 & 2018-06-09 20:54:00 & 0.776--0.810 & 3.9 & 5\,$\pm$\,1 & {\mybf 11\,$\pm$\,3} \\ 
00084747019 & 2018-06-17 15:25:00 & 2018-06-17 20:25:00 & 0.377--0.392 & 3.8 & 7\,$\pm$\,2 & {\mybf 16\,$\pm$\,4} \\ 
\enddata
\tablecomments{
$^a$ Phase zero is defined as the epoch of maximum flux in the \textsl{Fermi} LAT {\bfa (MJD 56498.7)}  . \\*
$^b$ The net exposure time spread over several snapshots.  Units are ks. \\*
$^c$ Count Rate is in the 2--10\,keV energy band.  Units are 10$^{-3}$\,counts s$^{-1}$. Errors are at the 1\,$\sigma$ level.\\* 
$^d$ Unabsorbed \textsl{Swift} XRT flux in the 1.0--10.0\,keV bandpass converted with PIMMS.  Units are 10$^{-13}$\,\ergcm2s.\\*
$^e$ 3 $\sigma$ upper limits.
}
\label{table:xrt}

\end{deluxetable}


\begin{deluxetable}{lccc}

\tablecaption{{\bfc 4FGL Spectral Parameters of Persistent High-Mass Gamma-ray Binaries}}
\tablehead{
\colhead{Source} & \colhead{$E_0$ (MeV)} & \colhead{$\alpha$ (lp index)} & \colhead{$\beta$}
}
\startdata
4FGL J1405.1-6119 & 1910.99 & 2.792 $\pm$ 0.076 & 0.302 $\pm$  0.048 \\
\tableline
4FGL J0240.5+6113$^a$  & 1177.76 & 2.396 $\pm$ 0.008 & 0.152 $\pm$ 0.005 \\
4FGL J0535.2-6736$^b$ & 752.28 & 2.609 $\pm$  0.090 & 0.153 $\pm$ 0.061 \\
4FGL J1018.9-5856$^c$ & 1772.77 & 2.678 $\pm$  0.023 & 0.236 $\pm$  0.014 \\
4FGL J1826.2-1450$^d$ & 1031.04 & 2.581 $\pm$ 0.023  & 0.124 $\pm$ 0.015 \\
\enddata
\tablecomments{
{\bfc
Spectral parameters are taken from the 4FGL LAT source catalog \citep{FermiLAT2019}. 
For all sources the \texttt{LogParabola} model was used, see Section \ref{sect:gray_results}.\\
$^a$ = \lsi\ (Be star) \\
$^b$ = LMC P3 (O star) \\
$^c$ = 1FGL J1018.6-5856 (O star) \\
$^d$ = LS 5039 (O star) \\
}
}
\label{table:hmgb_spectra}
\end{deluxetable}



\begin{deluxetable}{lccccccc}
\tablecolumns{8}
\tabletypesize{\small}
\tablewidth{0pc}
\tablecaption{Australia Telescope Compact Array Radio Measurements}
\tablehead{
\colhead{Time} & \colhead{Half Duration} &\colhead{Configuration}  &\colhead{Flux Density}
& \colhead{Error} & \colhead{Flux Density} & \colhead{Error}  \\
\colhead{(MJD)}  & \colhead{(days)} & \colhead{} &\colhead{5.5 GHz (mJy)}
& \colhead{5.5 GHz (mJy)} & \colhead{9 GHz (mJy)} & \colhead{9 GHz (mJy)}
}
\startdata
58074.051677 & 0.054630   &  1.5C  &  1.43 & 0.22 &  0.59 & 0.14 \\
58227.832580 & 0.034491   &  H168   & 0.35 & 0.09 &  0.36 & 0.09 \\
58258.472279 & 0.143056   &  6D   & 1.81 & 0.06 &  1.28 & 0.06 \\
58259.465913 & 0.148727   &  6D   & 1.49 & 0.11 &  1.25 & 0.08 \\
58266.432869 & 0.129572   &  6D   & 1.36 & 0.04 &  0.86 & 0.03 \\
58268.246122 & 0.064699   &  6D   & 0.85 & 0.22 &  0.51 & 0.08 \\
58277.279860 & 0.024132   &  6D   & 1.22 & 0.08 &  0.34 & 0.06 \\
58278.287325 & 0.023843   &  6D   & 1.32 & 0.07 &  0.70 & 0.06 \\
58409.903240 & 0.052257   &  6A   & 1.60 & 0.17 &  1.12 & 0.09 \\
58411.935300 & 0.049479   &  6A   & 1.03 & 0.40 &  1.15 & 0.17 \\
58427.816723 & 0.054514   &  6B   & 1.04 & 0.14 &  0.63 & 0.12 \\
58441.821527 & 0.003414   &  6B   & 1.31 & 0.26 &  0.77 & 0.11 \\
58445.044617 & 0.121181   &  6B   & 0.92 & 0.15 &  0.57 & 0.16 \\
58446.951851 & 0.091377   &  6B   & 0.82 & 0.05 &  0.53 & 0.05 \\
58460.747800 & 0.117766   &  H168   & 0.55 & 0.07 &  0.54 & 0.05 \\
\enddata
\tablecomments{The stated errors combine the statistical error, determined from RMS values in the region surrounding
\cxou, and a systematic error conservatively taken to be 5\% in the flux density scale between epochs.
The ATCA array configurations are the standard names for
the physical locations of the antennas: see
https://www.narrabri.atnf.csiro.au/operations/array\_configurations/configurations.html​
for full details of the antenna spacings in each array configuration.  
}
\label{table:atca}
\end{deluxetable}

\clearpage


\begin{deluxetable}{lcc}
\tablecolumns{8}
\tabletypesize{\small}
\tablewidth{0pc}
\tablecaption{Distance Determinations}
\tablehead{
\colhead{Color} & \colhead{R\&L} &\colhead{S\&F}  \\
\colhead{} & \colhead{distance (kpc) E(B-V)} &\colhead{distance (kpc) E(B-V)}
}
\startdata
H - K (2MASS) &  8.1 (8.76) & 6.6 (11.56) \\
H - K (VVV) & 7.9 (9.12) & 6.4 (12.0) \\
J - H (VVV) & 8.9 (8.65) & 8.0 (11.0)  \\

\enddata
\tablecomments{Distances in kpc determined using photometric
measurements from 2MASS \citep{Landi2017} and VVV \citep{Minniti2010},
the reddening laws
of \citet[][R\&L]{Rieke1985} and \citet[][S\&F]{Schlafly2011}, and the absolute magnitudes
of an O6.5 III star from \citet{Martins2006}.
}
\label{table:distance}
\end{deluxetable}

\clearpage

\end{document}